\documentclass[aps,prx,showpacs,superscriptaddress,nofootinbib,notitlepage,twocolumn]{revtex4-1}
\usepackage{graphicx}
\usepackage{color}
\usepackage{transparent}
\usepackage{amsmath,amssymb,amsfonts,amssymb}
\usepackage{epstopdf}
\usepackage{float}

\newcommand{\bra}[1]{\ensuremath{\left\langle#1\right|}}
\newcommand{\ket}[1]{\ensuremath{\left|#1\right\rangle}}
\newcommand{\bracket}[2]{\ensuremath{\left\langle#1 \vphantom{#2}\middle|  #2 \vphantom{#1}\right\rangle}}

\newcommand{\tr}[0]{\text{tr}}

\begin{document}

\title[]{Obtaining conclusive information from incomplete experimental quantum tomography}
\author{Henri Lyyra}
\affiliation{Turku Centre for Quantum Physics, Department of Physics and Astronomy, University of Turku, FIN-20014, Turun yliopisto, Finland}
\affiliation{QTF Centre of Excellence, Department of Physics and Astronomy, University of Turku, FI-20014 Turun Yliopisto, Finland}
\author{Tom Kuusela}
\affiliation{Turku Centre for Quantum Physics, Department of Physics and Astronomy, University of Turku, FIN-20014, Turun yliopisto, Finland}
\author{Teiko Heinosaari}
\affiliation{Turku Centre for Quantum Physics, Department of Physics and Astronomy, University of Turku, FIN-20014, Turun yliopisto, Finland}
\affiliation{QTF Centre of Excellence, Department of Physics and Astronomy, University of Turku, FI-20014 Turun Yliopisto, Finland}

\begin{abstract}
We demonstrate that incomplete quantum tomography can give conclusive information in experimental realizations. 
We divide the state space into a union of multiple disjoint subsets and determine conclusively which of the subsets a system, prepared in completely unknown state, belongs to. 
In other words we construct and solve membership problems. 
Our membership problems are partitions of the state space into a union of four disjoint sets formed by fixing two maximally entangled reference states and boundary values of fidelity function “radius” between the reference states and the unknown preparation. 
We study the necessary and sufficient conditions of the measurements which solve these membership problems conclusively. 
We construct and experimentally implement such informationally incomplete measurement on two-photon polarization states with combined one-qubit measurements, and solve the membership problem in example cases.
\end{abstract}

\maketitle

\section{Introduction}

Quantum tomography has become a standard procedure in quantum
information \cite{bisio2009}. Having a system prepared in an entirely unknown quantum state, it is possible to
identify the state by making suitably many measurements and reconstructing
it from the measurement data. The crucial point in quantum
tomography is that the overall collection of measurements is such that
each state gives unique measurement data. This kind of collection is
called informationally or tomographically complete \cite{busch1989}.

An incomplete collection of measurements does not allow a direct
state reconstruction from the measurement data as two different states
may give the same data. 
One can use some additional prior information, if available, to compensate the informational incompleteness. 
In particular, quantum tomography under the prior information that the unknown state is pure has been investigated in several earlier works \cite{weigert1992,finkelstein2004,flammia2005,heinosaari2013,chen2013,carmeli2015,kech2016,ma2016}.

In the current work we test a method in which the collection of
measurements is informationally incomplete and there is no prior information on the input state, but we still aim to obtain
conclusive information. 
Instead of trying to reconstruct the unknown state, we aim to decide from the measurement data in which of multiple disjoint subsets of states the unknown state belongs to. 
In recent theoretical works \cite{carmeli2014,carmeli2016,lu2016,carmeli2017} it has been
shown that whether this task is possible or not depends on the specific
details of the separation of the state space. For instance, deciding
whether an unknown state has rank greater than a boundary value $r$
or not is possible without a complete tomography if and only if $r$ is
smaller than $d / 2$, where $d$ is the dimension of the system \cite{carmeli2017}.

In the previously mentioned theoretical works several properties have
been identified whose verification requires complete tomography and
others that can be verified with some suitably chosen incomplete measurement setting. In the latter case, the actual implementation of the
procedure has not been studied at all. In particular, since the measurement setting is not complete, the usual state reconstruction methods
are not applicable.

In the current work, we report on our experiment of using incomplete
measurement setting to obtain conclusive information on the unknown
quantum state. The investigated system is the composite system of
two qubits, namely the total polarization state of two spatially separated photons. 
Since the photon pair is produced through a spontaneous parametric down-conversion process, the polarization system can be prepared in entangled polarization states. 
We solve the fidelity membership problem with respect to two maximally entangled reference states and the unknown polarization state by making only simultaneous local projective measurements on the photons, for two unknown state preparations. Restricting to combinations of local qubit projections prevents us from making projective measurements on the reference state, but still allows us to solve the membership problem. 

The general aim of this work is to demonstrate, by using data from an actual experiment constructed with the geometric tools developed in \cite{carmeli2016}, that one can obtain conclusive information even if the measurement setting is not informationally complete. 
The fact that we manage to solve the membership problem as predicted by our theoretical analysis is reflected by having conclusive decision on which of the partition segments contains the unknown state.
The confidence of the decision is so high that restrictions of numerical accuracy are more notable than the errors caused by imperfections of measurements. 
Further, we change the partition of the state space while keeping the initial state the same. This shows that the obtained information depends on the partition in the expected way. 

The paper is structured as follows.
We first identify, by applying the results from \cite{carmeli2017}, 
the fidelity membership problem that can be solved with incomplete tomography. 
We then provide a concrete measurement setting that is suitable for this task, and finally analyze the measurement data to obtain a conclusion. Our study demonstrates that the use of incomplete tomography to obtain
conclusive information is possible also in practice, not just in theory.

\section{Measurements and membership problems}

In quantum mechanics, any measurement can be described by a positive operator valued measure (POVM) $E$, which assigns a positive operator $E_j$ to each outcome $j$ of the measurement. When the system is prepared in state $\rho$, the probability of obtaining the outcome $j$ is $p(E_j) = \tr[E_j \rho]$. Thus, two states $\rho_1$ and $\rho_2$ can be distinguished by measurement $E$ if and only if measuring $E$ on both of them results to different probability distributions, or equivalently $\tr[E_j ( \rho_1 - \rho_2 )] \neq 0$ for some outcome $j$. 
 
If $E$ can distinguish any two states, it can be used in a tomographical measurement to fully construct the density matrix representation of the system's state. 
This type of measurements are called \textit{informationally complete measurements}. For a POVM $E$, we denote by dim$(E)$ the dimension of the linear span of the operators $E_j$, and we simply call this number the dimension of $E$. The informational completeness is then equivalent to dim$(E)=d^2$ \cite{busch1991}. If a measurement is not informationally complete, we say that it is \textit{informationally incomplete}.

\begin{figure}[H]
  \centering
  \includegraphics[width=.425\linewidth]{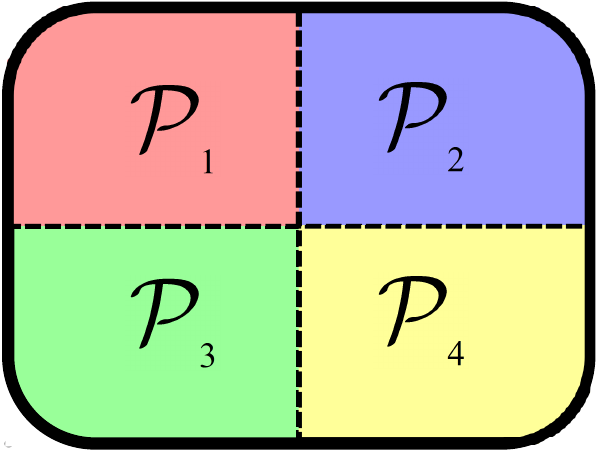}
    \caption{
    Illustration of a generic partition of the state space $\mathcal{S(H)}$ into a union of disjoint sets as $\mathcal{S(H)} = \mathcal{P}_1 \cup \mathcal{P}_2 \cup \mathcal{P}_3 \cup \mathcal{P}_4$. Here each $\mathcal{P}_k$ is represented by a segment of different color. In membership problems, the goal is to conclusively decide, which of the segments contains the unknown state.
    }
  \label{partition_gen}
\end{figure}

Let us consider a partition of the state space $\mathcal{S(H)}$ of Hilbert space $\mathcal{H}$ into disjoint subsets $\mathcal{P}_k$ of  $\mathcal{S(H)}$ such that $\mathcal{S(H)} = \cup_k \mathcal{P}_k$. We concentrate on problems of determining conclusively, which of the segments $\mathcal{P}_k$ contains the unknown state $\rho$ of the system. We call this kind of tasks \textit{membership problems} \cite{carmeli2017}.
A POVM $E$ can solve the membership problem if and only if it can distinguish every $\rho_1 \in \mathcal{P}_k$ from every $\rho_2 \in \mathcal{P}_n$, or equivalently if there does not exist a pair $\rho_1 \in \mathcal{P}_k$ and $\rho_2 \in \mathcal{P}_n$ with $k \neq n$ such that $\tr[E_j ( \rho_1 - \rho_2 )] = 0$ for all outcomes $j$. 

Since informationally complete measurements can distinguish any two states, trivially they can be used to solve any membership problem. Despite their versatility, using informationally complete measurements should be avoided in membership problems, if possible: since the number of parameters to be determined increases as $d^2$
, full tomography becomes experimentally very demanding and time-consuming for high-dimensional systems. 
It is also of foundational interest to understand which membership problems can be solved without an informationally complete measurement.

Since any membership problem can be formulated as a question -- \textit{in which of the sets $\mathcal{P}_k$ does the system state belong to?} -- it is tempting to assume that informationally complete measurements are not necessary to solve them. 
In \cite{carmeli2016,lu2016}, the possibility of solving the membership problem $\mathcal{S(H)} = \mathcal{P} \cup \mathcal{P}^C $ with informationally incomplete measurement was studied in cases, where $\mathcal{P}$ is the set of states of a bipartite quantum system whose subsystems share certain type of correlation. The results showed, that for some types of correlations, solving the membership problem requires informationally complete measurement, but for quantum discord it 
does not. 
Instead of the size of $\mathcal{P}$, the geometry of the boundary between $\mathcal{P}$ and $\mathcal{P}^C$ determines whether solving the membership problem requires informationally complete measurement or not.

Using the geometric framework introduced in \cite{carmeli2016}, other specific membership problems, namely norm distance, purity, rank, and fidelity membership problem, were studied in \cite{carmeli2017}. In this paper, we will concentrate on the fidelity membership problem and show how it can be solved experimentally in a two-photon polarization system with informationally incomplete measurement.

In the literature, there are two other common ways to gather information on a quantum state without the use of complete tomography. We briefly highlight the main differences to the current approach. First, in a \emph{state discrimination protocol} \cite{chefles2000} there are only a finite number of possible preparations and one tries to conclude the correct state from a finite number of measurement outcomes. In contrast, in the previously explained membership problem the state is completely unknown. The subsets $\mathcal{P}_k$ that define a membership problem typically all contain infinitely many states. Second, \emph{witnesses}, such as entanglement witnesses \cite{chsa2014}, are used to gain information on a certain property. However, a witness only corresponds to a sufficient criterion whereas in the membership problem we aim to have a conclusive decision. 


\section{Fidelity membership problem}

The fidelity $F(\rho,\sigma)$ between two states $\rho$ and $\sigma$ is defined as 
\begin{equation}
F(\rho,\sigma) := \tr \left[ \sqrt{ \sqrt{\rho} \sigma \sqrt{\rho} } \right]\,,
\end{equation}
where $\sqrt{ A }$ is the unique positive operator satisfying $\sqrt{ A }\sqrt{ A } = A$.
The fidelity satisfies $F(\rho,\sigma) = 1$ if and only if $\rho = \sigma$. Even though fidelity is not a proper metric, it is commonly used to quantify the closeness of quantum states \cite{QCQI}. 
In addition, it can be used to define a metric, namely the Bures distance as 
$D_B(\rho,\sigma) := \sqrt{ 2 - 2 F(\rho,\sigma) }\,,$ 
which is proportional to the Quantum Fisher information, an essential quantity in quantum metrology \cite{sommers2003,paris2009}.

By fixing a boundary state $\sigma$, we can use the fidelity between the \textit{reference state} $\sigma$ and the unknown state to form a membership problem. Now, the task is to determine, whether the unknown state is at least as close to $\sigma$ as some boundary value $\epsilon$ with respect to fidelity or not. In other words, we want to find in which part of the state space partition $\mathcal{S(H)} = \mathcal{S}^{\ge \epsilon}_{\sigma} \cup \mathcal{S}^{< \epsilon}_{\sigma} $ the unknown state is. Here, 
we have denoted
\begin{align}\label{partition_fid}
\begin{aligned}
\mathcal{S}^{\ge \epsilon}_{\sigma} &= \{ \rho \in \mathcal{S(H)} : F( \rho, \sigma) \ge \epsilon \}\,,~~ \text{and}	\\ 
\mathcal{S}^{< \epsilon}_{\sigma} &= \{ \rho \in \mathcal{S(H)} : F( \rho, \sigma) < \epsilon \}
\end{aligned}
\end{align}
for any $0 \le \epsilon \le 1$. Let $E$ be a POVM with the set of so-called perturbation operators $\Delta$, defined as 
\begin{equation}
\mathcal{X}_E := \{\Delta \in \mathcal{L(H)} : \tr[\Delta E_j] = 0 ~ \forall j,\, \tr[\Delta] = 0,\, \Delta^* = \Delta \},
\end{equation} 
where $\Delta^*$ denotes the Hermitian conjugate of $\Delta$, and $\mathcal{L(H)}$ is the space of linear operators on $\mathcal{H}$. In \cite{carmeli2017}, it was shown that the fidelity membership problem can be solved conclusively by measuring the POVM $E$ with perturbations $\Delta$ satisfying 
\begin{equation}\label{deltacond}
\sqrt{\sigma} \Delta \sqrt{\sigma} = 0\,.
\end{equation}
Here, solving the membership problem conclusively means that for any state $\rho \in \mathcal{S(H)}$ the measurement outcome distribution can be analyzed in such a way that it tells us that $\rho$ belongs to either $\mathcal{S}^{\ge \epsilon}_{\sigma}$ or $\mathcal{S}^{< \epsilon}_{\sigma}$, but never in both of them.

If the condition of Eq.~\eqref{deltacond} is violated by a perturbation $\Delta \in \mathcal{X}_{E'}$ of POVM $E'$, there exists at least one pair of states $\rho_1 \in \mathcal{S}^{\ge \epsilon}_{\sigma}$ and $\rho_2 \in \mathcal{S}^{< \epsilon}_{\sigma}$ such that measuring $E'$ on them results to exactly the same outcome probability distribution $\tr[E'_j \rho_1] = \tr[E'_j \rho_2]\,\forall\,j$. Thus, analyzing the measurement data of $E'$ when the system was prepared to state $\rho$ which is one of these states would lead to \textit{inconclusive solution of the membership problem}: the system was prepared in a state $\rho$ which can equally likely belong to either $\mathcal{S}^{\ge \epsilon}_{\sigma}$ or $\mathcal{S}^{< \epsilon}_{\sigma}$, since measuring $E'$ on $\rho_1$ and $\rho_2$ leads to the exact same distribution.

In \cite{carmeli2017}, an upper bound for the minimal dimension of the POVM $E$, able to solve the fidelity membership problem, was shown to be dim$(E) = r^2 + 1$. Here $r$ is the rank of the reference state $\sigma$. For pure state $\sigma = \ket{\varphi_\sigma}\bra{\varphi_\sigma}$, we get $r = 1$, and thus dim$(E) = 2$, so there exists a POVM with only two outcomes, such that it conclusively solves our membership problem. 
The elements of such a  POVM can be chosen as $E_1 = \sigma,\, E_2 = \mathbb {I} - \sigma$. This can be verified from the condition \eqref{deltacond}, which reads now 
\begin{equation}
\ket{\varphi_\sigma}\bra{\varphi_\sigma} \Delta \ket{\varphi_\sigma}\bra{\varphi_\sigma} = \bra{\varphi_\sigma} \Delta \ket{\varphi_\sigma}\ket{\varphi_\sigma}\bra{\varphi_\sigma} = 0 \, , 
\end{equation}
and hence reduces to $\bra{\varphi_\sigma} \Delta \ket{\varphi_\sigma} = 0$. 
 Now, by expanding $\ket{\varphi_\sigma}$ to an orthonormal basis $\mathcal{B}$ of $\mathcal{H}$, the definition of the perturbations gives us 
\begin{align}
\begin{aligned}
0 	& = \tr[\Delta E_1]\\
 	& = \tr[\Delta \ket{\varphi_\sigma}\bra{\varphi_\sigma}]\\
 	& = \sum_{\psi \in \mathcal{B}} \bra{\psi}\Delta \ket{\varphi_\sigma}\bracket{\varphi_\sigma}{\psi}\\
 	& = \bra{\varphi_\sigma}\Delta \ket{\varphi_\sigma}\,,
\end{aligned}
\end{align}
which shows that each $\Delta \in \mathcal{X}_E$ satisfies Eq.~\eqref{deltacond}. 

The number of segments in the membership problem can be increased by using simultaneously multiple reference states. For example, using two reference states, $\chi$ and $\xi$, and fixing their corresponding fidelity boundary values, $\alpha$ and $\beta$, we can form the four segmented partition of the state space:
\begin{equation}\label{membership4_0}
\mathcal{S(H)} = \mathcal{P}_1 \cup \mathcal{P}_2 \cup \mathcal{P}_3 \cup \mathcal{P}_4\,,
\end{equation}
where
\begin{align}
\begin{aligned}
\mathcal{P}_1 &= \mathcal{S}^{< \alpha}_{\chi} \cap \mathcal{S}^{\ge \beta}_{\xi}\,, ~
\mathcal{P}_2 = \mathcal{S}^{\ge \alpha}_{\chi} \cap \mathcal{S}^{\ge \beta}_{\xi}\,, ~\\
\mathcal{P}_3 &= \mathcal{S}^{< \alpha}_{\chi} \cap \mathcal{S}^{< \beta}_{\xi}\,,~  
\mathcal{P}_4 = \mathcal{S}^{\ge \alpha}_{\chi} \cap \mathcal{S}^{< \beta}_{\xi}\,.
\end{aligned}
\end{align}
In order to solve the extended membership problem in Eq.~\eqref{membership4_0}, the POVM $E$ now has to satisfy the condition of Eq.~\eqref{deltacond} for both $\sigma = \chi$ and $\sigma = \xi$.

For multi-partite systems, the projections on all pure states cannot be performed with simultaneous local measurements on the subsystems when the total system state is entangled. In what follows, we show how the fidelity membership problem with respect to maximally entangled states can be solved with informationally incomplete simultaneous local projections on the subsystems of a two-qubit photonic system.
 
 From now on, we restrict to the two-qubit cases when $\sigma$ is one of the Bell states $\ket{\Psi^-} = \frac{1}{\sqrt{2}} ( \ket{0,1} - \ket{1,0} )$ or $\ket{\Psi^+} = \frac{1}{\sqrt{2}} ( \ket{0,1} + \ket{1,0} )$. 
We fix the matrix representation $\ket{0} = (1~0)^\text{T}, \ket{1} = (0~1)^\text{T}$ and denote $\Delta$ with the generic Hermitian two-qubit operator
\begin{equation}
\Delta =
\left(
\begin{array}{cccc}
 a_{1,1} & a_{1,2} e^{i \theta _{1,2}} & a_{1,3} e^{i \theta _{1,3}} & a_{1,4} e^{i \theta _{1,4}} \\
 a_{1,2} e^{-i \theta _{1,2}} & a_{2,2} & a_{2,3} e^{i \theta _{2,3}} & a_{2,4} e^{i \theta _{2,4}} \\
 a_{1,3} e^{-i \theta _{1,3}} & a_{2,3} e^{-i \theta _{2,3}} & a_{3,3} & a_{3,4} e^{i \theta _{3,4}} \\
 a_{1,4} e^{-i \theta _{1,4}} & a_{2,4} e^{-i \theta _{2,4}} & a_{3,4} e^{-i \theta _{3,4}} & a_{4,4} \\
\end{array}
\right)
\,.
\end{equation}
By using this in Eq.~\eqref{deltacond}, we see that the fidelity membership problem with respect to $\Psi^-$ and $\Psi^+$ is solved by measuring $E$ if and only if all $\Delta \in \mathcal{X}_E$ satisfy $ a_{2,2} + a_{3,3} -2 a_{2,3} \cos \left(\theta _{2,3} \right)  = 0$ and $ a_{2,2} + a_{3,3} + 2 a_{2,3} \cos \left(\theta _{2,3} \right) = 0$, respectively. As a consequence, measuring $E$ solves simultaneously the fidelity membership problem with respect to $\Psi^-$ and $\Psi^+$ if and only if 
\begin{align}\label{deltacondexpl}
a_{2,2} &= -a_{3,3} ~~ \text{and} ~~ \Re \left( a_{2,3} e^{i \theta_{2,3}} \right) = 0\,. 
\end{align}

By using the two reference states, $\Psi^-$ and $\Psi^+$, and their corresponding bipartite partitions as defined by Eq.~\eqref{partition_fid}, we form the following four segmented partition of the state space
\begin{equation}\label{membership4}
\mathcal{S(H)} = \mathcal{P}_1 \cup \mathcal{P}_2 \cup \mathcal{P}_3 \cup \mathcal{P}_4\,,
\end{equation}
where
\begin{align}
\begin{aligned}
\mathcal{P}_1 &= \mathcal{S}^{< \epsilon^-}_{\Psi^-} \cap \mathcal{S}^{\ge \epsilon^+}_{\Psi^+}\,, ~
\mathcal{P}_2 = \mathcal{S}^{\ge \epsilon^-}_{\Psi^-} \cap \mathcal{S}^{\ge \epsilon^+}_{\Psi^+}\,, ~\\
\mathcal{P}_3 &= \mathcal{S}^{< \epsilon^-}_{\Psi^-} \cap \mathcal{S}^{< \epsilon^+}_{\Psi^+}\,,~
\mathcal{P}_4 = \mathcal{S}^{\ge \epsilon^-}_{\Psi^-} \cap \mathcal{S}^{< \epsilon^+}_{\Psi^+}\,.
\end{aligned}
\end{align}
Next, we will present our experimental setup and show how to implement an informationally incomplete measurement $E$ which solves the membership problem of Eq.~\eqref{membership4}.


\section{The experiment}

\subsection{The experimental setup}

\begin{figure}[H]
  \centering
  \includegraphics[width=1\linewidth]{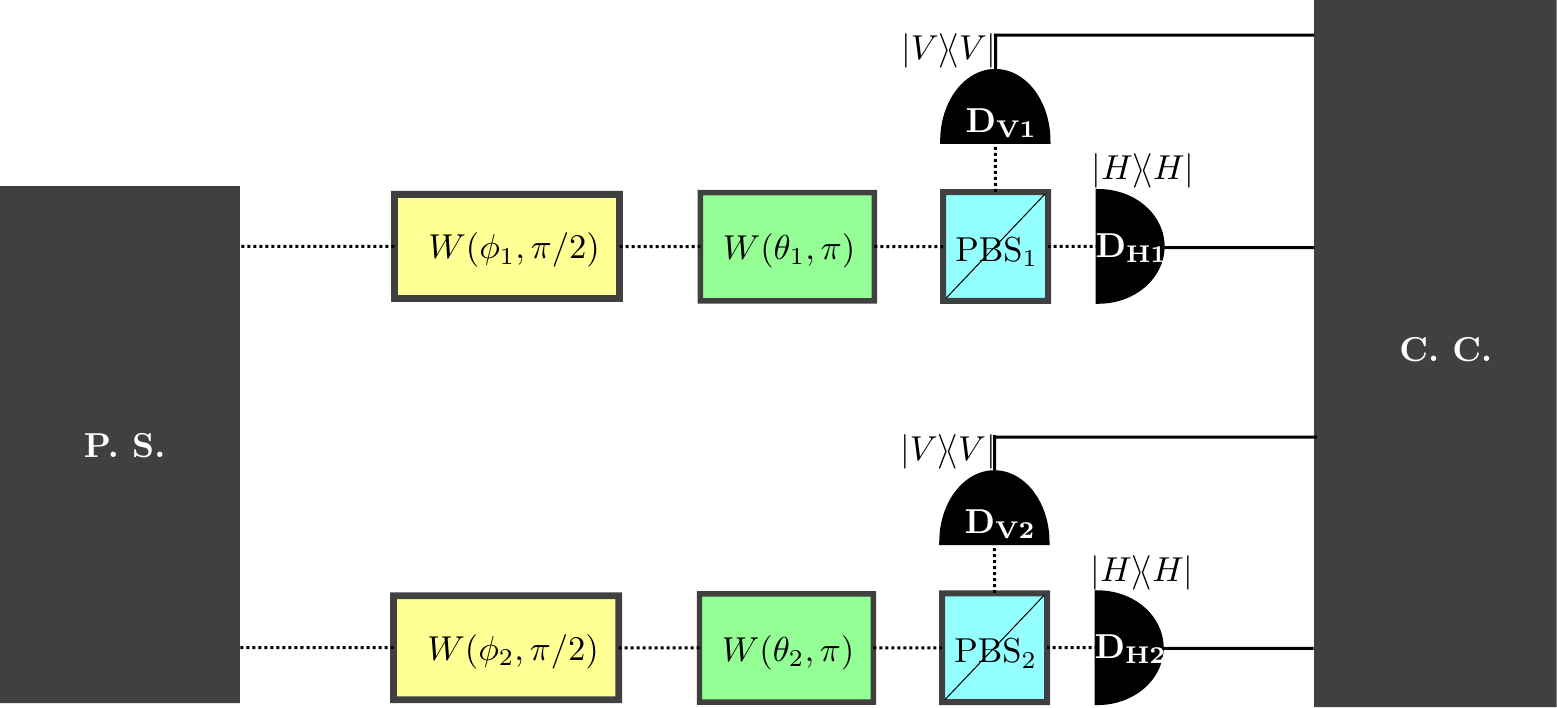}
    \caption{Schematic illustration of the experimental setup. P.~S. the photon source which produces the unknown 2 photon polarization states, $W(\phi_k,\pi/2)$ quarter-wave plates, $W(\theta_k,\pi)$ half-wave plates, PBS$_k$ polarizing beam splitters, $D_{ik}$ single-photon detectors for $k\in\{1,\,2\}$ and $i\in\{H,\,V\}$, and C.~C. the coincidence counting electronics which collects the statistics of the coincidences of photon detections.}
  \label{detection}
\end{figure}

From now on, we concentrate on a specific quantum optical system, namely the polarization of two photons. We define the matrix representation of polarization through the $\{0,\,1\}$ basis as $\ket{H} := \ket{0},\,\ket{V} := \ket{1} $, where $H$ ($V$) corresponds to horizontal (vertical) polarization. 
In the experiment, a 2 mm thick type II beta-barium-borate crystal is pumped with 40 mW single-mode continuous wave laser diode operating at 405 nm. The spontaneous parametric down-conversion process in crystal produces a pair of photons in polarization entangled state $\ket{\Psi^-} = \frac{1}{\sqrt{2}} ( \ket{H,V} - \ket{V,H} )$. We label the photons as $1$ and $2$. From the source, the photons pass through interference filters with 10 nm full width at half maximum centered at 810 nm. Then, the photons are coupled to single mode optical fibers and guided into their respective detection stations, illustrated in Fig.~\ref{detection}.


In the detection stations, the projective measurements on the polarization qubits are manipulated by rotating half-wave plates $W(\theta_k,\pi)$ and quarter-wave plates $W(\phi_k,\pi/2)$, where $k\in\{1,\,2\}$. 
%
In the fixed matrix representation, the action of a wave plate on the one-photon polarization state $\rho$ can be written as $W(\mu,\,\nu) \rho W(\mu,\,\nu)^*$, where
\begin{align}
W(\mu,\,\nu) = 
\begin{pmatrix}
\cos^2(\mu) + e^{i \nu}\sin^2(\mu)		&	\frac{1}{2}(1-e^{i \nu})\sin(2\mu)			\\
\frac{1}{2}(1-e^{-i \nu})\sin(2\mu)		&	\sin^2(\mu)	 + e^{i \nu}\cos^2(\mu)
\end{pmatrix}
\end{align}
and $\mu$ and $\nu$ correspond to rotation angle and phase shift of the wave plate, respectively. In the detection stations of Fig.~\ref{detection}, this means that the total two-qubit measurement bases can be rotated with operators 
\begin{align}\nonumber
&A_{1,2}(\theta_1,\phi_1,\theta_2,\phi_2):=A_1(\theta_1,\phi_1)\otimes A_2(\theta_2,\phi_2),\, \text{where} \\
&A_k(\theta_k,\phi_k) := W(\phi_k,\,\pi/2)^* W(\theta_k,\,\pi)^*,\, \text{and}\, k\in\{1,\,2\}\,. 
\end{align}

After the wave plates, each photon goes through a polarizing beam splitter PBS$_k$ and ends up at a detector $D_{Vk}$ or $D_{Hk}$. 
For each rotated polarization basis, the measurement data consists of coincidence counts in detector combinations 
\begin{align}
\{(H1,\,H2),\,(H1,\,V2),\,(V1,\,H2),\,(V1,\,V2)\}\,,
\end{align}
  whose measurement outcome probabilities in the polarization basis rotated with angles $\theta_1,\phi_1,\theta_2,$ and $\phi_2$ are obtained as 
\begin{align}
&p(i,\,j,\,\theta_1,\phi_1,\theta_2,\phi_2) =\text{tr}\big[ P_{i,j}(\theta_1,\phi_1,\theta_2,\phi_2) \rho\big]\,,
\end{align}
where $i,\,j\in\{H,\,V\}$, and the total polarization system is prepared in state $\rho$. 
For short, we denote the projections on the rotated basis elements, corresponding to our POVM elements, as
\begin{equation}
\begin{split}
& P_{i,j}(\theta_1,\phi_1,\theta_2,\phi_2) \\ \label{bases_gen}
&:= A_{1,2}(\theta_1,\phi_1,\theta_2,\phi_2) \ket{i}\hspace{-0.125cm}\bra{i}\otimes\ket{j}\hspace{-0.125cm}\bra{j} A^{*}_{1,2}(\theta_1,\phi_1,\theta_2,\phi_2)\,,
\end{split}
\end{equation}
where $i,\,j\in\{H,\,V\}$.

\subsection{POVM to solve the membership problem of fidelity environments}

\begin{table}[H]
\centering
\caption{
Three orthonormal bases, $\mathcal{B}_1,\,\mathcal{B}_2$ and $\mathcal{B}_3$, forming a sufficient set of projective measurements to solve the fidelity membership problem of two-photon polarization states with respect to $\Psi^-$ and $\Psi^+$. Here $\theta_k$ corresponds to rotation angle of half-wave plate and $\phi_k$ corresponds to rotation of quarter-wave plate of photon $k \in \{ 1,\,2 \}$. 
}
\label{bases1}
\begin{tabular}{l|ccc}
\hline\hline
           		& ~$\mathcal{B}_1$~ 		& ~$\mathcal{B}_2$~		& ~$\mathcal{B}_3$~\\ \hline
~$\phi_1$~   	& ~~~0~~~       		& ~~$\pi/4$~~			& ~~0~~ \\
~$\theta_1$~ 	& ~~~0~~~      			& ~~$\pi/8$~~			& ~~$\pi/8$~~ \\ 
~$\phi_2$~   	& ~~~0~~~     			& ~~$\pi/4$~~			& ~~0~~ \\
~$\theta_2$~ 	& ~~~0~~~       		& ~~$\pi/8$~~			& ~~$\pi/8$~~ \\ \hline\hline
\end{tabular}
\end{table}

In Table \ref{bases1}, we present three combinations of wave plate rotation angles, corresponding to measurements of different orthonormal bases (ONB), $\mathcal{B}_1,\,\mathcal{B}_2$ and $\mathcal{B}_3$, used in the experiment. $\mathcal{B}_1$ consists of the tensor products of projections on local $\{H,\,V\}$ bases, $\mathcal{B}_2$ consists of the tensor products of projections on local $\{+,\,-\}$ bases, and $\mathcal{B}_3$ consists of the tensor products of projections on local $\{R,\,L\}$ bases. Thus, the three bases are mutually unbiased. 

For each basis, the measurement is repeated for multiple identical copies of the unknown state and the probability distribution of the measurement outcomes is collected. In the measurement, the total set of POVM elements is $E = \frac{1}{3}\mathcal{B}_1 \cup \frac{1}{3}\mathcal{B}_2 \cup \frac{1}{3}\mathcal{B}_3$, where the ONB's, defined by Eq.~\eqref{bases_gen} with the rotation angle combinations of Table \ref{bases1}, become 
\begin{align}\nonumber
\mathcal{B}_1 =& 
\left\{
\begin{pmatrix}
 1 & 0 & 0 & 0 \\
 0 & 0 & 0 & 0 \\
 0 & 0 & 0 & 0 \\
 0 & 0 & 0 & 0 
\end{pmatrix},\,
\begin{pmatrix}
 0 & 0 & 0 & 0 \\
 0 & 1 & 0 & 0 \\
 0 & 0 & 0 & 0 \\
 0 & 0 & 0 & 0 
\end{pmatrix},\,
\right.
\\ & \left. ~\,
\begin{pmatrix}
 0 & 0 & 0 & 0 \\
 0 & 0 & 0 & 0 \\
 0 & 0 & 1 & 0 \\
 0 & 0 & 0 & 0 
\end{pmatrix},\,
\begin{pmatrix}
 0 & 0 & 0 & 0 \\
 0 & 0 & 0 & 0 \\
 0 & 0 & 0 & 0 \\
 0 & 0 & 0 & 1 
\end{pmatrix}
\right\}
\\ \nonumber
\mathcal{B}_2 =& 
\left\{
\frac{1}{4}
\begin{pmatrix}
 1 & 1 & 1 & 1 \\
 1 & 1 & 1 & 1 \\
 1 & 1 & 1 & 1 \\
 1 & 1 & 1 & 1 
\end{pmatrix},\,
\frac{1}{4}
\begin{pmatrix}
 1 & -1 & 1 & -1 \\
 -1 & 1 & -1 & 1 \\
 1 & -1 & 1 & -1 \\
 -1 & 1 & -1 & 1 
\end{pmatrix},\,
\right.
\\ & \left. ~\,
\frac{1}{4}
\begin{pmatrix}
 1 & 1 & -1 & -1 \\
 1 & 1 & -1 & -1 \\
 -1 & -1 & 1 & 1 \\
 -1 & -1 & 1 & 1 
\end{pmatrix},\,
\frac{1}{4}
\begin{pmatrix}
 1 & -1 & -1 & 1 \\
 -1 & 1 & 1 & -1 \\
 -1 & 1 & 1 & -1 \\
 1 & -1 & -1 & 1 
\end{pmatrix}
\right\}
\\ \nonumber
\mathcal{B}_3 =& 
\left\{
\frac{1}{4}
\begin{pmatrix}
 1 & i & i & -1 \\
 -i & 1 & 1 & i \\
 -i & 1 & 1 & i \\
 -1 & -i & -i & 1 
\end{pmatrix},\,
\frac{1}{4}
\begin{pmatrix}
 1 & -i & i & 1 \\
 i & 1 & -1 & i \\
 -i & -1 & 1 & -i \\
 1 & -i & i & 1 
\end{pmatrix},\,
\right.
\\ & \left. ~\,
\frac{1}{4}
\begin{pmatrix}
 1 & i & -i & 1 \\
 -i & 1 & -1 & -i \\
 i & -1 & 1 & i \\
 1 & i & -i & 1
\end{pmatrix},\,
\frac{1}{4}
\begin{pmatrix}
 1 & -i & -i & -1 \\
 i & 1 & 1 & -i \\
 i & 1 & 1 & -i \\
 -1 & i & i & 1
\end{pmatrix}
\right\}
\end{align}
We number the POVM elements $E_i$ so that $E_{j + 4(k-1)}$ is the $j^\text{th}$ element of $\mathcal{B}_k$ scaled with factor 1/3. Since the dimension of this POVM is 10, and the POVM is informationally complete if and only if dim$(E) = d^2 = 16$, we conclude that $E$ is informationally incomplete.

By solving a basis for the kernel of the linear space spanned by $E$, we find the space of the perturbations to be
\begin{align}\nonumber
\mathcal{X}_E = \text{Span}
&\left( \left\{ 
\begin{pmatrix}
 0 & i & 0 & 0 \\
 -i & 0 & 0 & 0 \\
 0 & 0 & 0 & -i \\
 0 & 0 & i & 0 
\end{pmatrix},\,
\begin{pmatrix}
 0 & 0 & i & 0 \\
 0 & 0 & 0 & -i \\
 -i & 0 & 0 & 0 \\
 0 & i & 0 & 0 
\end{pmatrix},\,
\right. \right.
\\
& ~~~~~
\begin{pmatrix}
 0 & 0 & 0 & 0 \\
 0 & 0 & -i & 0 \\
 0 & i & 0 & 0 \\
 0 & 0 & 0 & 0 
\end{pmatrix},\,
\begin{pmatrix}
 0 & 0 & 0 & -i \\
 0 & 0 & 0 & 0 \\
 0 & 0 & 0 & 0 \\
 i & 0 & 0 & 0 
\end{pmatrix},\,
\\ \nonumber
&\left. \left. ~~~~
\begin{pmatrix}
 0 & -1 & 0 & 0 \\
 -1 & 0 & 0 & 0 \\
 0 & 0 & 0 & 1 \\
 0 & 0 & 1 & 0 
\end{pmatrix},\,
\begin{pmatrix}
 0 & 0 & -1 & 0 \\
 0 & 0 & 0 & 1 \\
 -1 & 0 & 0 & 0 \\
 0 & 1 & 0 & 0 
\end{pmatrix}
  \right\} \right)
\end{align}

Each of the basis elements of $\mathcal{X}_E$ satisfies the conditions in Eq.~\eqref{deltacondexpl} 
and, as a consequence, so does every $\Delta \in \mathcal{X}_E$. This shows that measuring the POVM $E$ solves the fidelity membership problems with respect to the reference states $\Psi^-$ and $\Psi^+$, as described above. Since $E$ solves the membership problem with respect to both of the reference states simultaneously, it also solves the four segmented membership problem of Eq.~\eqref{membership4}.
In order to solve the membership problem, we analyze the measurement outcome probability distributions of all the three bases. 

\begin{table}[H]
\centering
\caption{
Nine orthonormal bases, $\mathcal{B}_1,\,\mathcal{B}_2$ and $\mathcal{B}_3',\dots,\mathcal{B}_9'$, forming an insufficient set of projective measurements to solve the fidelity membership problem of two-photon polarization states with respect to $\Psi^-$ and $\Psi^+$. Here $\theta_k$ corresponds to rotation angle of half-wave plate and $\phi_k$ corresponds to rotation of quarter-wave plate of photon $k \in \{ 1,\,2 \}$. Note that the bases $\mathcal{B}_1$ and $\mathcal{B}_2$ are the same as in Table \ref{bases1}, but instead of $\mathcal{B}_3$ this set includes bases $\mathcal{B}_3',\dots,\mathcal{B}_9'$.
}
\label{basesbad}
\begin{tabular}{l|ccccccccc}
\hline
\hline
           		& ~\,$\mathcal{B}_1$\,~ 	& ~\,$\mathcal{B}_2$\,~ 	& ~\,$\mathcal{B}_3'$~\, 	& ~\,$\mathcal{B}_4'$~\, 	& ~\,$\mathcal{B}_5'$~\, 	& ~\,$\mathcal{B}_6'$~\, 	& ~\,$\mathcal{B}_7'$~\, 	&~\, $\mathcal{B}_8'$~\, 	& ~\,$\mathcal{B}_9'$\,~ \\ \hline
~$\phi_1$~   	& $0$    		& $\pi/4$     	& $\pi/4$	& $\pi/8$  	& $\pi/8$    	& $\pi/8$	& $\pi/8$     	& $\pi/8$     	& $\pi/8$ \\
~$\theta_1$~ 	& $0$     	& $\pi/8$     	& $0$		& $0$     	& $0$       	& $\pi/4$	& $\pi/4$    	& $\pi/4$    	& $\pi/4$ \\ 
~$\phi_2$~   	& $0$     	& $\pi/4$     	& $\pi/4$	& $\pi/4$     	& $0$       	& $0$		& $\pi/4$     	& $\pi/8$     	& $\pi/8$ \\ 
~$\theta_2$~ 	& $0$     	& $\pi/8$     	& $0$		& $0$     	& $0$       	& $0$		& $0$     	& $0$       	& $\pi/4$ \\ \hline\hline
\end{tabular}
\end{table}

For the sake of example, we present in Table \ref{basesbad} a set of nine ONB's resulting to another POVM. Here, the bases $\mathcal{B}_1$ and $\mathcal{B}_2$ are exactly the same as in Table \ref{bases1} and the basis $\mathcal{B}_3$ has been replaced by 7 other bases, namely $\mathcal{B}_3',\dots,\mathcal{B}_9'$. Similarly to the bases of Table \ref{bases1}, we can use the projective measurements on the basis elements of $\mathcal{B}_1,\mathcal{B}_2,\mathcal{B}_3',\dots,\mathcal{B}_9'$ of Table \ref{basesbad} and form another POVM $E'$. Using $E'$, we find a basis for its space of perturbations $\mathcal{X}_{E'}$. Checking the condition in Eq.~\eqref{deltacond} shows that these measurements cannot be used to solve conclusively our membership problem since there exists at least one $\Delta' \in \mathcal{X}_{E'}$ such that Eq.~\eqref{deltacond} is violated. This means that there exist states for which the measurement outcome distribution of measuring $E'$ could be produced by two states in different segments of the partition, leading to inconclusive result. 

It is worth noting, that the dimension of the POVM $E$ in Table \ref{bases1} is 10 whereas the dimension of the POVM $E'$ in Table \ref{basesbad} is 13. This serves as an example of how higher dimension of the POVM does not necessarily mean that it is more capable of solving the membership problem.

\subsection{Measurement results}\label{results}

The membership problem was solved for two unknown states, Preparation 1 and Preparation 2, in the experiment. For the measurement in each basis, the average 
coincidence rate was 400 per second and the measurement time 60 seconds. Thus, the contribution of multiphoton events was negligible. Any unbalance in the beam splitters and losses in the collecting optics of optical fibers were compensated, as also differences in the quantum efficiencies of the single photon detectors. The dark count rate of the detectors was less than 200 counts/s and the width of the coincidence time window was 10 ns.

Preparation 1 (2) was prepared close to the state $\Psi^-$ ($\Psi^+$). In Table \ref{measdata1} and \ref{measdata2}, we present the measured outcome probabilities in each basis. Here, the outcome probabilities are evaluated by their relative frequencies $p(P_{i,j}) = C(i,j)/(\sum_{i,j}C(i,j))$, where $C(i,j)$ is the number of coincidence counts of the projection outcomes $(i1,j2)$ and $i,j \in \{H,\,V\}$.

\begin{table}[H]
\centering
\caption{
Measurement bases and the corresponding measured coincidence count probabilities of the detectors $H1, V1, H2,$ and $V2$ in Fig. 2, for the unknown state Preparation 1. Here, the rotation angles $\theta_1,\phi_1,\theta_2$ and $\phi_2$ are specified by the bases $\mathcal{B}_1,\mathcal{B}_2$, and $\mathcal{B}_3$ as presented in Table 1.
}
\label{measdata1}
\begin{tabular}{c|cccc}
\hline\hline
        			& $p(P_{HH})$ & $p(P_{HV})$ & $p(P_{VH})$ & $p(P_{VV})$ \\ \hline
~$\mathcal{B}_1$~ 	& ~ 0.0295 ~  & ~ 0.4541 ~  & ~ 0.4836 ~  & ~ 0.0328 ~   \\ 
~$\mathcal{B}_2$~ & ~ 0.0798 ~  & ~ 0.3974 ~  & ~ 0.4459 ~  & ~ 0.0769 ~  \\ 
~$\mathcal{B}_3$~ & ~ 0.0827 ~ & ~ 0.3998  ~ & ~ 0.4341 ~  & ~ 0.0834  ~ \\ \hline\hline
\end{tabular}
\end{table}

\begin{table}[H]
\centering
\caption{
Measurement bases and the corresponding measured coincidence count probabilities of the detectors $H1, V1, H2,$ and $V2$ in Fig. 2, for the unknown state Preparation 2.
}
\label{measdata2}
\begin{tabular}{c|cccc}
\hline\hline
        & $p(P_{HH})$ & $p(P_{HV})$ & $p(P_{VH})$ & $~ p(P_{VV})$~ \\ \hline
~$\mathcal{B}_1$~ &~  0.026305~  & ~ 0.443038~  &~  0.499604~  &~  0.031052~  \\ 
~$\mathcal{B}_2$~ & ~ 0.446745~  &~  0.054748~  &~  0.075055~  &~  0.423452~  \\ 
~$\mathcal{B}_3$~ & ~ 0.45232 ~  & ~ 0.057764 ~ &~  0.054239~  &~  0.435677~  \\ \hline\hline
\end{tabular}
\end{table}

\begin{figure*}[t!]
\begin{minipage}[t]{1\textwidth}
\includegraphics[width=1\textwidth]{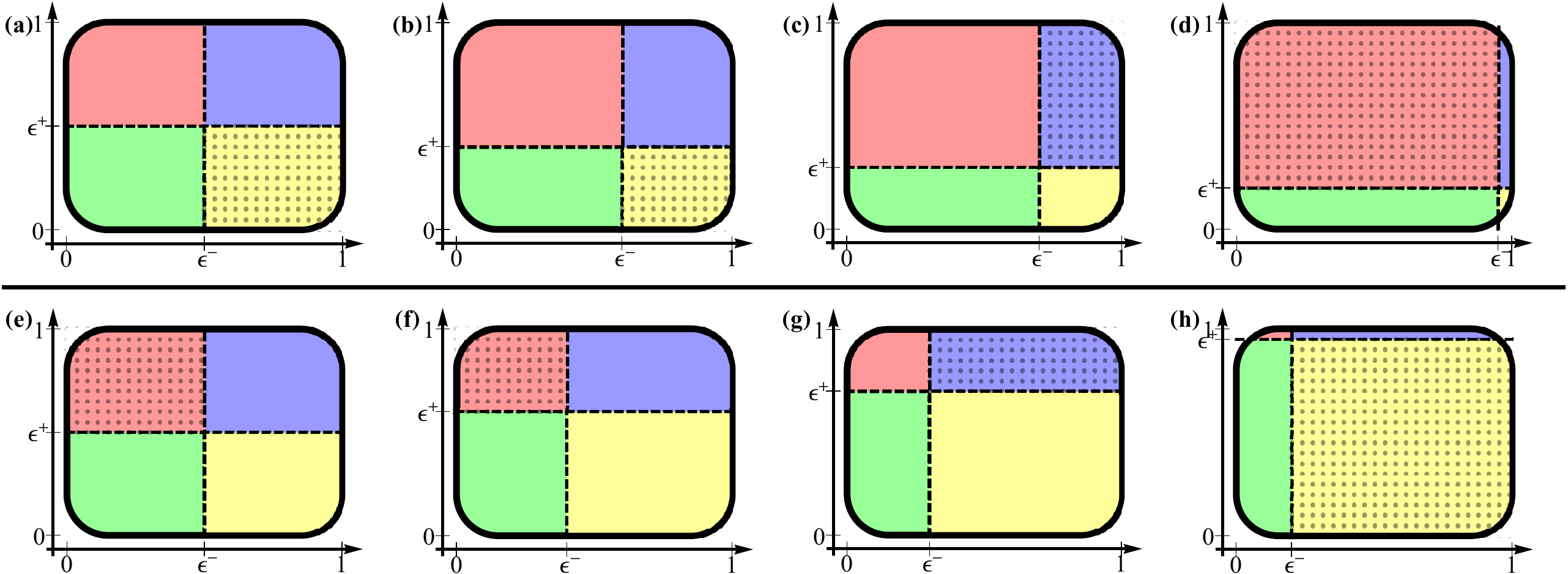}
    \caption{
    \textbf{The analysis of the membership problems }
    \textbf{(a) -- (d):} for Preparation 1. \textbf{(e) -- (h):} for Preparation 2. For both unknown state preparations, the boundary fidelity values, $\epsilon^-$ and  $\epsilon^+$, with respect to the reference states $\Psi^-$ and $\Psi^+$, are changed in each panel to change the partition $\mathcal{S(H)} = \mathcal{P}_1 \cup \mathcal{P}_2 \cup \mathcal{P}_3 \cup \mathcal{P}_4$. Once the values of $\epsilon^-$ and  $\epsilon^+$ are fixed, partition of the state space into four disjoint segments is formed. Here, $\epsilon^-$ and  $\epsilon^+$ are visualized with the vertical and horizontal dashed lines, respectively, and the disjoint segments are illustrated with different colors. In each segment, the density operator $\rho(\textbf{b})$ which produces the measurement outcome distribution closest to the experimentally measured one is found numerically, as described in Sec.~\ref{results}. In each partition, we find that in exactly one of the segments the optimal density operator $\rho(\textbf{b})$ produces the same outcome distribution as measured. Thus, we know conclusively which segment contains the unknown state and mark it by dotting. 
    The $\epsilon$ values in each panel are: 
    \textbf{(a)} $\epsilon^- = 0.5,$ $\epsilon^+ = 0.5,\,$\textbf{(b)} $\epsilon^- = 0.6,$ $\epsilon^+ = 0.4,\,$\textbf{(c)} $\epsilon^- = 0.7,$ $\epsilon^+ = 0.3,$  and \textbf{(d)} $\epsilon^- = 0.95,$ $\epsilon^+ = 0.2$.
     \textbf{(e)} $\epsilon^- = 0.5,$ $\epsilon^+ = 0.5,\,$\textbf{(f)} $\epsilon^- = 0.4,$ $\epsilon^+ = 0.6,\,$\textbf{(g)} $\epsilon^- = 0.3,$ $\epsilon^+ = 0.7,\,$\textbf{(h)} $\epsilon^- = 0.2,$ $\epsilon^+ = 0.95$.
    }
  \label{partition_sol_2}
\end{minipage}
\end{figure*}

The measured probability distributions of the three bases are combined into a normalized vector $\textbf{p}$, resulting to a single probability distribution. 
We use the SLSQP optimization algorithm in the Python method \texttt{scipy.optimize.minimize} to numerically solve the density matrix $\rho(\textbf{b})$, which produces the measurement outcome distributions closest to the measured distributions. The density matrix is parametrized with the generalized Bloch vector $\textbf{b} \in \mathbb{R}^{15}$ as $\rho(\textbf{b}) = \frac{1}{4} \mathbb{I} + \textbf{b} \cdot \mathbf{\Gamma}$, where the $b_i = \tr [ \Gamma_i \rho(\textbf{b})]$ and $\Gamma_i$ are the 4$\times$4 generalizations of Gell-Mann matrices \cite{bertlmann}. 

We use $\textbf{b}$ as the optimization parameter and minimize the $\ell_1$-distance $d_1(\textbf{p},\textbf{q}) = \sum_{i = 1}^{N} \vert p_i - q_i \vert$, where $p_i$ is the probability of an outcome $i$, corresponding to the POVM element $E_i$, in the experiment, $q_i = \tr[ E_i  \rho(\textbf{b}) ]$, and $N$ is the number of elements of the POVM. 
  The positivity of $\rho(\textbf{b})$ is guaranteed by using the positivity of its eigenvalues as optimization constraint. Applying the additional constraint $\rho(\textbf{b}) \in \mathcal{S}^{\ge \epsilon}_{\sigma}$ in SLSQP guarantees that $d_1(\textbf{p},\textbf{q}) = 0$ if and only if $\rho \in \mathcal{S}^{\ge \epsilon}_{\sigma}$ for the unknown state $\rho$ prepared in the experiment.

Since we solve the optimization problem numerically, the $\ell_1$-distance can never be exactly zero due to the limitations of numerical precision.  We conclude that $\ell_1$-distance is zero whenever $d_1(\textbf{p},\textbf{q}) \approx 10^{-8}$. This is in line with the numerical accuracy when using simulated error-free measurement data. In cases where $d_1(\textbf{p},\textbf{q}) \approx 10^{-8}$ for the optimal states $\rho(\textbf{b}_1)$ and $\rho(\textbf{b}_2)$ in two different segments $\mathcal{P}_1$ and $\mathcal{P}_2$, we conclude that the unknown state is so close to the boundary between $\mathcal{P}_1$ and $\mathcal{P}_2$ that the numerical and experimental errors are large enough to make the result inconclusive. In such cases, we have to change the boundary values $\epsilon^-$ and  $\epsilon^+$ to change the partition so that we can get conclusive result.

Note that the density operator $\rho(\textbf{b})$ is not necessarily the unknown state, even if the $\ell_1$-distance  is zero. This is because, for informationally incomplete measurements, two different states can lead to the same measurement outcome distributions. Nevertheless, since our measurement $E$ is constructed to distinguish any state of $\mathcal{S}_\sigma^{<\epsilon}$ from any state of $\mathcal{S}_\sigma^{\ge\epsilon}$ (for $\sigma = \Psi^-$ and $\sigma = \Psi^+$), the $\ell_1$-distance between the measured outcome distribution and the outcome distribution of the optimized (non-unique) density operator $\rho(\textbf{b})$ is zero if and only if the unknown state belongs to the same segment of the partition as $\rho(\textbf{b})$. Thus, if we get $d_1(\textbf{p},\textbf{q}) = 0$ for $\rho(\textbf{b})$ in $\mathcal{S}_\sigma^{<\epsilon}$, we know that also the unknown state belongs to $\mathcal{S}_\sigma^{<\epsilon}$ and similarly for $\mathcal{S}_\sigma^{\ge\epsilon}$.

First the analysis is performed to the unknown state Preparation 1. The fidelity boundary value $\epsilon^-$ is fixed, and the optimization is performed for the reference state $\sigma = \Psi^-$ to solve the membership problem $\mathcal{S(H)} = \mathcal{S}^{\ge \epsilon^-}_{\Psi^-} \cup \mathcal{S}^{< \epsilon^-}_{\Psi^-}$. Then $\epsilon^+$ is fixed and the optimization is performed for the reference state $\sigma = \Psi^+$, solving the membership problem $\mathcal{S(H)} = \mathcal{S}^{\ge \epsilon^+}_{\Psi^+} \cup \mathcal{S}^{< \epsilon^+}_{\Psi^+}$. This is repeated for multiple choices of $\epsilon^-$ and $\epsilon^+$, corresponding to different partitions. Then the unknown state is changed to Preparation 2 and the protocol is performed again with different choices of $\epsilon^-$ and $\epsilon^+$. We collect the results for both preparations and different values of $\epsilon^-$ and $\epsilon^+$ in Fig.~\ref{partition_sol_2}.

In Fig.~\ref{partition_sol_2}, we present the results for two unknown state preparations in the experiment. For both preparations, we show the solution to the fidelity membership problem for four different partitions. The difference between the partitions is the values of the fidelity boundaries $\epsilon^-$ and $\epsilon^+$, illustrated by the vertical and horizontal dashed lines, respectively. The area to the right (left) from the vertical dashed line at $\epsilon^-$, is the set of states whose fidelity with $\Psi^-$ is at least $\epsilon^-$ (smaller than $\epsilon^-$). The area above (below) the horizontal dashed line at $\epsilon^+$, is the set of states whose fidelity with $\Psi^+$ is at least $\epsilon^+$ (smaller than $\epsilon^+$). The dotted area denotes the segment of the partition which contains the unknown state.

In panels \textbf{(a) -- (d)} of Fig.~\ref{partition_sol_2}, we show the solutions to four different fidelity membership problems when the system is in the unknown state Preparation 1. In panel \textbf{(a)}, where the boundary values are set as $\epsilon^- = 0.5,\, \epsilon^+ = 0.5$, we see that the unknown state belongs to the bottom-right corner of the partition. When in panel $\textbf{(b)}$ the values are changed to $\epsilon^- = 0.6,\, \epsilon^+ = 0.4$, the partition is changed, but the unknown state is still in the bottom-right corner. Setting values to  $\epsilon^- = 0.7,\, \epsilon^+ = 0.3 $ further changes the partition and the unknown state belongs to the set on the top-right corner as shown in \textbf{(c)}. In \textbf{(d)} we see how the partition changes as we set $\epsilon^- = 0.95,\, \epsilon^+ = 0.2$ and the unknown state is contained by the top-left set of this new partition.

In panels \textbf{(e) -- (h)} of Fig.~\ref{partition_sol_2}, we show the results for the unknown state Preparation 2. We see in panel \textbf{(e)}, that for $\epsilon^- = 0.5,\, \epsilon^+ = 0.5$ , the unknown state belongs to the top-left corner. This is in contrast to what happened for Preparation 1 in panel \textbf{(a)} for the same boundary values. In \textbf{(f)}, the partition is changed by setting $\epsilon^- = 0.4,\, \epsilon^+ = 0.6$. We see that the unknown state still belongs to the top-left segment of the partition. Choosing the values as $\epsilon^- = 0.3,\, \epsilon^+ = 0.7$ changes the partition so that the unknown state is contained by the top-right segment this time. Finally, in panel \textbf{(h)}, we see that changing the partition by choosing $\epsilon^- = 0.2,\, \epsilon^+ = 0.95$ causes the unknown state to belong to the bottom-right segment.

In this proof-of-principle paper, we concentrated on the case of two reference states resulting to four segmented membership problems. In principle, the number of segments in the partition can be increased by using more reference states. For example, choices $\sigma = \ket{\varphi_\sigma}\bra{\varphi_\sigma}$, where $\ket{\varphi_\sigma} = \ket{0,0},$ $\ket{\varphi_\sigma} = \ket{0,1},$ $\ket{\varphi_\sigma}= \ket{1,0},$ $\ket{\varphi_\sigma} = \ket{1,1},$ $\ket{\varphi_\sigma} = \frac{1}{\sqrt{2}}( \ket{0,0} - \ket{1,1} ),$ and $\ket{\varphi_\sigma} = \frac{1}{\sqrt{2}}( \ket{0,0} + \ket{1,1} ),$ also satisfy the condition of Eq.~\eqref{deltacond} for the perturbations of our POVM $E$. This shows that $E$ can also solve conclusively the bipartite fidelity membership problems with respect to all of these reference states. Using these with $\Psi^-$ and $\Psi^+$ as reference states and their corresponding fidelity boundary values $\epsilon^k$ generalizes the partition in Eq.~\eqref{membership4} to cover a situation of 8 reference states. This leads to $(\epsilon^1,\,\epsilon^2,\dots,\epsilon^8)$-parametrized family of membership problems with $2^8$ disjoint segments. This way, even more complicated $(\epsilon^{1},\dots,\epsilon^n)$-parametrized $2^n$ segmented membership problems for $n$ reference states can be constructed. 

We conclude that in each partition and for both state preparations in the experiment, we find that in exactly one of the segments the optimal density operator $\rho(\textbf{b})$ produces the same outcome distribution as measured. In other words, we have found out conclusively in each case which segment contains the unknown state and thus we have solved the membership problems. Since our POVM $E$ was constructed so that it satisfies the condition of Eq.~\eqref{deltacond}, this serves as experimental evidence for the validity of the theoretical geometric tools of \cite{carmeli2016} for constructing informationally incomplete measurements which solve membership problems.

\section{Conclusions}

We have constructed a family of $(\epsilon^-, \, \epsilon^+)$-parametrized four segmented membership problems in the two-qubit state space. The partition was formed by fixing two maximally entangled two-qubit states, namely $\Psi^-$ and $\Psi^+$ as example reference states and using the dividing boundary values $\epsilon^-$ and $\epsilon^+$ of the fidelity between the reference states and the unknown state. Using the theoretical results of \cite{carmeli2016,carmeli2017}, we have studied the necessary and sufficient conditions of the POVM which can solve these membership problems. 

We have constructed an informationally incomplete POVM capable of solving these membership problems and experimentally implemented it in the optical setup of two-photon polarization states by restricting to simultaneous local projective measurements. We illustrated the problem by using two unknown state preparations in the experiment and four pairs of $(\epsilon^-,\,\epsilon^+)$ to form different membership problems for each preparation in the analysis. We have shown how to numerically analyze the measurement results to solve the membership problems. As our analysis shows, the unknown state was found to belong to exactly one for the segments, and thus the membership problems were conclusively solved in each case. 
Even though our work concentrates on parametrized four segmented membership problems, we have shown that measuring our POVM $E$ solves also a $8^2$ segmented parametrized membership problem formed with 8 different reference states. This way we illustrated how the geometric tools can be used to solve complicated $(\epsilon^{1},\dots,\epsilon^n)$-parametrized $2^n$ segmented membership problems of $n$ reference states.

To conclude, we have presented a successful experimental test for the recently developed geometric tools, presented in \cite{carmeli2016}, for solving quantum membership problems with informationally incomplete measurements. We wish that our proof-of-principle experiment paves the way for informationally incomplete experimental implementations of other geometrically approachable membership problems in the future, such as the quantum discord and rank problems.\\ 

\section*{Acknowledgements}

H.L.~acknowledges the financial support from the University of Turku Graduate School (UTUGS). T.H.~acknowledges financial support from the Academy of Finland via the Centre of Excellence program (Project no.~312058) as well as Project no.~287750.

\section*{Author contribution statement}
T.H.~proposed the original idea. Most of theoretical analysis was performed by H.L. under the supervision of T.H. T.K.~implemented the experiments. The paper was written by H.L., T.K., and T.H..

\end{document}